\documentclass[10pt,aps,prd,twocolumn,showpacs,amsmath,amssymb,nofootinbib,preprintnumbers]{revtex4-1}
\pdfoutput=1

\usepackage{graphicx}
\usepackage{subfigure}
\usepackage{amssymb}
\usepackage{slashed}
\usepackage{hyperref}
\usepackage{dsfont}
\hypersetup{colorlinks=true,citecolor=blue}

\usepackage{tikz}
\usepackage{calc}

\newcommand{\cA}{\ensuremath{\mathcal A}}
\newcommand{\cB}{\ensuremath{\mathcal B}}
\newcommand{\cC}{\ensuremath{\mathcal C}}
\newcommand{\cG}{\ensuremath{\mathcal G}}
\DeclareMathOperator{\vol}{vol}
\def\Re{{\rm Re \,}}
\def\Im{{\rm Im \,}}

\def\ZZ{{\mathbb Z}}
\def\RR{{\mathbb R}}

\def\sm{\smallskip}

\begin{document}
  \title{Holographic duals for five-dimensional superconformal quantum field theories}

 \author{Eric D'Hoker}
 \email{dhoker@physics.ucla.edu}
 \author{Michael Gutperle}
 \email{gutperle@physics.ucla.edu}
 \author{Christoph F.~Uhlemann}
 \email{uhlemann@physics.ucla.edu}

 \affiliation{Mani L.\ Bhaumik Institute for Theoretical Physics\\
Department of Physics and Astronomy\\
University of California, Los Angeles, CA 90095, USA \\ \vskip 0in
{\sl Dedicated to John H. Schwarz on the occasion of his seventy fifth birthday.}}

\begin{abstract}
We construct global solutions to Type IIB supergravity with 16 residual supersymmetries whose space-time is  $AdS_6 \times S^2$ warped over a Riemann surface. Families of solutions are labeled by an arbitrary number $L\geq 3$ of asymptotic regions, in each of which the supergravity fields match those of a $(p,q)$ five-brane, and may therefore be viewed as near-horizon limits of fully localized intersections of  five-branes in Type IIB string theory. These solutions provide compelling candidates for holographic duals to a large class of five-dimensional superconformal quantum field theories which arise as non-trivial UV fixed points of perturbatively non-renormalizable Yang-Mills theories, thereby making them more directly accessible to quantitative analysis.
\end{abstract}

\maketitle

\section{Introduction}

Yang-Mills theories in dimensions greater than four are perturbatively non-renormalizable.
Nevertheless, investigations of Coulomb branch dynamics  in supersymmetric gauge theory, and of brane configurations in string theory, provide convincing evidence that some five-dimensional supersymmetric Yang-Mills theories can flow to non-trivial ultraviolet  fixed points \cite{Seiberg:1996bd,Morrison:1996xf,Intriligator:1997pq}. The resulting superconformal field theories (SCFT) are strongly interacting, exhibit a variety of global symmetries such as the exceptional Lie group $E_8$, and do not possess a standard Lagrangian description. 
Aside from the intrinsic  interest these mysterious theories present, a further motivation for their study is a certain similarity to Einstein gravity in four dimensions which is also perturbatively non-renormalizable but for which a non-trivial fixed point, if it exists, would lead to a well-defined theory.

\sm

\begin{figure}
\begin{center}
\subfigure[][]{\label{fig:weba}
\includegraphics[]{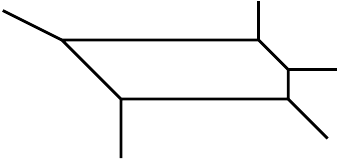}
}
\qquad\qquad
\subfigure[][]{\label{fig:webb}
\includegraphics[]{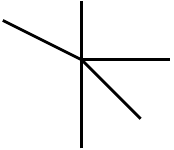}
}
\end{center}
\caption{
Fig.~\ref{fig:weba} shows a $(p,q)$ 5-brane web for an SU(2) Yang-Mills theory with one flavor at a generic point on the Coulomb branch. Fig.~\ref{fig:webb} shows the counterpart of (a) for the conformal limit at the origin of the Coulomb branch. It is characterized by the $(p,q)$ charges of the external 5-branes, which give the slopes of the branes in the figure. }
\label{fig:web}
\end{figure}

A natural realization of Coulomb branch dynamics  in five-dimensional supersymmetric Yang-Mills theory is provided by webs of intersecting five-branes in Type IIB string theory \cite{Aharony:1997ju,Aharony:1997bh}. Five-branes transform  under the $SL(2,\ZZ)$ duality group of Type IIB  \cite{Schwarz:1995dk}, and may  be labelled by two co-prime integers $(p,q)$, with NS5 and D5-branes carrying the charges $(1,0)$ and $(0,1)$. The requirements of charge conservation and 16 residual supersymmetries constrain the angles spanned between the branes at every intersection in the web. The superconformal limit of a brane web is obtained by letting the finite lengths of branes tend to zero, thereby collapsing the web to a star-shaped figure, as illustrated in fig.~\ref{fig:web}.

\sm

Neither the dynamics on the Coulomb branch in supersymmetric gauge theory, nor the structure of $(p,q)$ five-brane webs in string theory give us, however, direct quantitative access to the superconformal theories. Holographic methods, which are applicable when the gauge group has large rank,  provide an ideal tool to fill this gap. Five-dimensional conformal $SO(2,5)$ symmetry requires  supergravity dual solutions built on $AdS_6$. A sparse set of singular $AdS_6$ solutions to Type IIA supergravity was found in \cite{Brandhuber:1999np,Bergman:2012kr},  and evidence for their relation to five-dimensional SCFTs was presented in \cite{Jafferis:2012iv,Alday:2014rxa}. The T-duals give solutions in Type IIB which exhibit yet further singularities \cite{Lozano:2013oma}, and no $AdS_6$ solutions directly in Type IIB have been obtained. 

\sm

In this letter we construct all global solutions with 16 supersymmetries in Type IIB supergravity, whose space-time is  $AdS_6 \times S^2$ warped over a Riemann surface $\Sigma$ with the topology of a disc, using the local solutions derived recently in  \cite{D'Hoker:2016rdq}. Families of global solutions are characterized by an arbitrary number $L \geq 3$ of asymptotic regions in which the supergravity fields coincide with  those of a $(p,q)$ five-brane. We shall argue that these solutions provide holographic duals for five-dimensional superconformal field theories, and thereby open the way to quantitative studies using the tools of AdS/CFT.

\sm

The symmetries and the Ansatz for supergravity fields are reviewed in sec.~\ref{sec:iib-ansatz}, while the local solutions of  \cite{D'Hoker:2016rdq} are summarized in sec.~\ref{sec:local-sol}. Global supergravity solutions are constructed in sec.~\ref{sec:global-sol}, and related to five-brane webs describing five-dimensional SCFTs in sec.~\ref{sec:brane-webs}. We close with a discussion in sec.~\ref{sec:discussion}.

\section{Type IIB supergravity Ansatz }
\label{sec:iib-ansatz}

The superconformal algebra in five dimensions is unique and given by the  Lie superalgebra $F(4)$, whose maximal bosonic subalgebra $SO(2,5) \oplus SO(3)$ is dictated by conformal symmetry and R-symmetry \cite{Nahm:1977tg, Minwalla:1997ka}. Its 16 fermionic generators distinguish it from counterparts in dimensions 3, 4, and 6 where the number of fermionic generators of the maximal superconformal algebra is 32.

\sm

The  fields of Type IIB supergravity are the metric $g_{MN}$, the complex axion-dilaton scalar $B$, the complex 2-form $C_{(2)}$, the real 4-form $C_{(4)}$, as well as the gravitino and the dilatino fields \cite{Schwarz:1983qr,Howe:1983sra}.

In the supergravity description, the bosonic symmetry $SO(2,5) \oplus SO(3)$ is realized as an isometry and the full space-time is given by $(AdS_6 \times S^2) \ltimes \Sigma$, where the product is warped over a Riemann surface $\Sigma$.
The symmetry requirement similarly restricts the fluxes and leads to the following Ansatz for the 
Type IIB supergravity fields,
\begin{align}
\label{1}
\begin{split}
 ds^2 &= f_6^2 \, ds^2 _{AdS_6} + f_2 ^2 \, ds^2 _{S^2} + ds^2 _\Sigma~,
 \\
 C_{(2)}&=\cC \vol_{S^2}~,
 \qquad\qquad
 C_{(4)}=0~,
\end{split}
\end{align}
where $ds^2 _{AdS_6}$ and $ds^2 _{S^2}$ are the metrics of unit radius invariant under $SO(2,5)$ and $SO(3)$ respectively, while  $\vol_{S^2}$ is the volume form on S$^2$. We may choose local complex coordinates $w,\bar w$ in which the metric on $\Sigma$  is conformally flat and given by $ds^2 _\Sigma = 4 \rho^2 |dw|^2$.  The metric factors $f_6, f_2, \rho$, and the fields $B$ and $\cC$ then depend only on $\Sigma$, while the gravitino and dilatino fields vanish.

\section{Local solutions}
\label{sec:local-sol}

Type IIB supergravity solutions invariant under $SO(2,5) \oplus SO(3)$ and 16 supersymmetries are obtained by reducing the BPS equations to the supergravity fields of the Ansatz (\ref{1}), and then solving these reduced BPS equations. Several earlier attempts to follow this strategy  fell short of solving the reduced BPS equations \cite{Apruzzi:2014qva,Kim:2015hya,Kim:2016rhs}. 

\sm

Recently, the BPS equations were integrated completely in \cite{D'Hoker:2016rdq}, by  building on the methods developed in \cite{D'Hoker:2007xy}. The intermediate manipulations are quite involved, but the final result is simple and may be expressed in terms of two locally holomorphic functions  $\cA_\pm$ on the Riemann surface $\Sigma$. The solution may be expressed  with the help of the variables $\kappa ^2, \cG$, and $R$ defined by, 
\begin{align}
\label{eq:W-R-def}
\nonumber
 \kappa^2&=-|\partial_w \cA_+|^2+|\partial_w \cA_-|^2~,
 \\
  \nonumber
 \partial_w\cB&=\cA_+\partial_w \cA_- - \cA_-\partial_w\cA_+~,
 \\
 \nonumber
 \cG&=|\cA_+|^2-|\cA_-|^2+\cB+\bar{\cB}~,
 \\
  R+\frac{1}{R}&=2+6\,\frac{\kappa^2 \, \cG }{|\partial_w\cG|^2}~.
\end{align}
In terms of these variables, and an additional integration constant $c_6$, the supergravity fields of the solutions can be expressed concisely. The metric functions read
\begin{align}
\begin{split}
 f_2^2&=\frac{c_6^2  \kappa^2 (1-R)}{9 \, \rho^2 (1+R)}~,
 \qquad
 f_6^2=\frac{c_6^2 \kappa^2 (1+R) }{\rho^2 \, (1-R)}~,
 \\
 \rho^2&=\frac{ c_6 (R+R^2)^\frac{1}{2} }{|\partial_w \cG|} \left (  \frac{\kappa ^2 }{1-R} \right )^{\frac{3}{2}}~,
\end{split}
\end{align}
the axion-dilaton field $B$ is given by 
\begin{align}
B &=\frac{\partial_w \cA_+ \,  \partial_{\bar w} \cG - R \, \partial_{\bar w} \bar \cA_-   \partial_w \cG}{
R \, \partial_{\bar w}  \bar \cA_+ \partial_w \cG - \partial_w \cA_- \partial_{\bar w}  \cG}~,
\end{align}
while the flux potential $\cC$ takes the following form
\begin{align}\label{eqn:flux}
 \cC &= \frac{4 i c_6}{9}\left [  \frac{\partial_{\bar w}\bar\cA_- (R^2+1) \partial_w\cG
 -2R\partial_w\cA_+\partial_{\bar w} \cG}{(R+1)^2 \, \kappa^2 } \right .
\nonumber \\ & \hskip 1in \left . - \bar  \cA_- - 2 \cA_+\right ]~.
\end{align}
In the local solutions presented above, the locally holomorphic functions $\cA_\pm$ are arbitrary. 
The constant $c_6$ can be absorbed into a rescaling of $\cA_\pm$ and we will therefore set it to one in the following.

\section{Global solutions}
\label{sec:global-sol}

To obtain physically well-defined global solutions, the supergravity fields must satisfy the reality, positivity, and regularity conditions derived in \cite{D'Hoker:2016rdq}. For example, reality and positivity of the metric functions $f_6^2, f_2^2, \rho^2$ requires  $R$ of (\ref{eq:W-R-def}) to be real and  $\kappa ^2, \cG$, and $(1-R)$ to have the same sign. Furthermore, $\Sigma$ necessarily needs to have a non-empty boundary  \cite{ads6-2}. The full set of reality, positivity and regularity conditions is given by  
\begin{align}
\label{8}
 \kappa^2&>0~,& \cG&>0
\end{align}
in the interior of $\Sigma$, along with  the boundary conditions
\begin{align}
\label{9}
 \kappa^2\big\vert_{\partial\Sigma}=\cG\big\vert_{\partial\Sigma}&=0~.
\end{align}
These conditions ensure that the sphere $S^2$ degenerates to form a regular $S^3$ sphere while $AdS_6$ retains a non-vanishing radius, such that the full ten-dimensional geometry is smooth and does not have a boundary. The main result of this letter, to be derived in the remainder of this section, is the solution for the locally holomorphic functions $\cA_\pm$ subject to the conditions (\ref{8}) and (\ref{9}).

\sm

The first step in the resolution of this problem hinges on an  analogy to two-dimensional electrostatics \cite{ads6-2} which we  summarize below. Since $\cA_\pm$ are locally holomorphic, the ratio $\lambda = \partial_w\cA_+/\partial_w\cA_-$ is locally meromorphic, and the real function $\Phi= -\ln |\lambda|^2$ is harmonic where $\lambda$ is holomorphic.  The function $\Phi$ may be interpreted as an electrostatics potential. The regularity conditions on $\kappa^2$ imply that $\Phi$  needs to be strictly positive in the interior of $\Sigma$ and must vanish on the boundary $ \partial \Sigma$.

\sm

In this letter, we shall restrict attention to the simplest case when $\Sigma$ has genus zero and $\partial\Sigma$ has one component, so that $\Sigma$ has the topology of  the upper half plane with $\partial \Sigma=\mathds{R}$. Assuming that $\partial_w \cA_\pm$ are meromorphic throughout $\Sigma$,  the function $\Phi$ is constructed as the potential for an arbitrary distribution of positive charges in the upper half plane along with negative mirror charges in the lower half plane,  thereby automatically ensuring  $\Phi >0$ in the interior of $\Sigma$ and $\Phi=0$ on $\partial\Sigma$.

\sm

\begin{figure}
\includegraphics[]{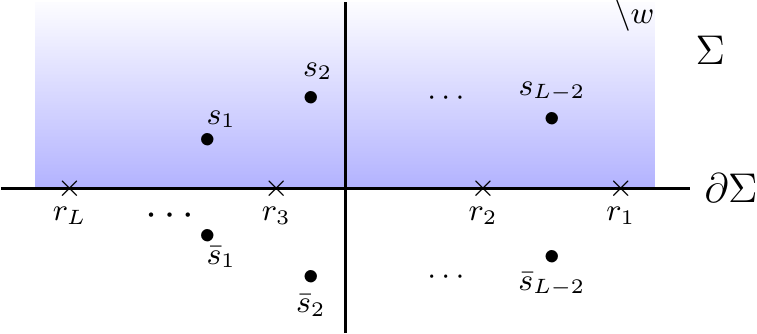}
\caption{The upper half plane $\Sigma$
with boundary $\partial \Sigma = \mathds{R}$.
Zeros of the function $\lambda$  are at the points $s_n$ and poles at $\bar s_n$, 
with $n=1,\cdots, L-2$ and $\Im (s_n) >0$.
The differentials $\partial_w \cA_\pm$ have poles at the 
points $r_\ell \in \mathds{R}$ with $\ell =1, \cdots, L$.
\label{fig1}}
\end{figure}

The function $\lambda$ is constructed by holomorphically splitting the potential $\Phi$. To obtain a single-valued $\lambda$ we require unit charges, so that $\lambda$ is a rational function with simple zeros at $s_n$ and simple poles at $\bar s_n$ with $\Im (s_n) >0$ for $n=1,\cdots, L-2$. Regularity of the  differentials $ \partial _w \cA_\pm$ requires them to have  $L$ common poles $r_\ell$ on the real axis for $\ell = 1, \cdots, L$, as illustrated in  fig.~\ref{fig1}. Integrating $\partial_w \cA_\pm$ gives the functions $\cA_\pm$  as follows, 
\begin{align}
 \cA_\pm (w) &=\cA_\pm^0+\sum_{\ell=1}^L Z_\pm^\ell \ln(w-r_\ell)~,
\end{align}
where the residues obey $Z_\pm^\ell= - \overline{Z_\mp^\ell}$ and are  given by
\begin{align}
 Z_+^\ell
 &=
 \sigma\prod_{n=1}^{L-2}(r_\ell-s_n)\prod_{k \neq\ell}^L\frac{1}{r_\ell-r_k}~,
\end{align}
while $\cA^0_\pm$ and $\sigma$ are complex constants. The construction guarantees $\kappa^2>0$ in the interior of $\Sigma$ and $\kappa^2=0$  on $\partial \Sigma$.
Furthermore, due to $\sum_{\ell} Z_\pm^\ell=0$ the functions $\cA_\pm$ are regular at infinity
and $\Sigma$ can be mapped to a disc.

\sm

Next, we enforce the boundary condition of (\ref{9}) on $\cG$. The analysis involves  careful consideration of the logarithmic branch cuts, and gives $\cG$ as follows, 
\begin{align}
\label{11}
\begin{split}
 \cG&=
\sum _{\ell < k}^L  Z^{[\ell k]}  \Bigg  \{
\left (  \ln \frac{w-r_k}{(r_\ell-r_k)^2} \right )\, \left ( \,  \overline{\ln \frac{ w-r_\ell}{(r_\ell-r_k)^2} } \, \right )
\\ &\hphantom{=}
+ \int _ \infty ^w dz \, \left ( \frac{ \ln (z -r_\ell)}{z - r_k } -  \frac{ \ln (z -r_k)}{z - r_\ell } \right )
- \hbox{c.c}
\Bigg \}\,,
\end{split}
\end{align}
where $Z^{[\ell k]}\equiv Z_+^\ell Z_-^k-Z_+^k Z_-^\ell$. The branch cuts of the logarithms are chosen to be on the real axis to the left so that, for $x$ real and positive,  $\ln (x)$ is real  and $\ln (e^{i \pi} x)=\ln(x)+i\pi$. When $w$ is real and to the right of all poles $r_\ell$, inspection of (\ref{11}) readily reveals that $\cG=0$ there. To guarantee that $\cG$ remains zero as $w$ crosses each pole $r_k$ in turn, we must implement one condition per pole on the parameters $s_n, r_\ell, \cA_\pm ^0, \sigma$, which takes the form
\begin{align}
\label{eqn:constr}
 \cA^0 Z_-^k + \bar \cA^0 Z_+^k 
+ \sum _{\ell \not= k }Z^{[\ell k]} \ln |r_\ell - r_k| &=0~.
\end{align}
Here, we have set $2\cA^0=\cA^0_+-\bar\cA_-^0$, and note that  the combination $2\cA_+^0 + \bar\cA_-^0$ effects a gauge transformation on $\cC$ and therefore parametrizes physically equivalent solutions.
The sum over all conditions in (\ref{eqn:constr}) vanishes, since $\sum_\ell Z_\pm^\ell=0$ by construction,
so only $L-1$ conditions are independent. 

\sm

Finally, we enforce the condition (\ref{8}) that $\cG >0$ in the interior of $\Sigma$. Instead of obtaining this result directly from (\ref{11}), we use a short indirect argument, based on the  relation $\partial_w\partial_{\bar w}\cG=-\kappa^2$ between $\kappa ^2$ and $\cG$.  Using this equation,  the vanishing of $\cG$  on $\partial \Sigma$, the positivity of the scalar Green function on $\Sigma$ with Dirichlet conditions on $\partial \Sigma$, and the positivity of $\kappa^2$ in the interior of $\Sigma$, it readily follows that $\cG>0$ in the interior of $\Sigma$.

\sm

The functions $\cA_\pm$ constructed above directly provide, via (\ref{eq:W-R-def})--(\ref{eqn:flux}), explicit Type IIB supergravity solutions. These solutions are manifestly  regular throughout $\Sigma$ and its boundary, except perhaps at the isolated points corresponding to the poles $r_\ell$  where a separate analysis is required which will be presented in the next section.

\section{Connection to five-brane webs}
\label{sec:brane-webs}

In this section, we shall show that the Type IIB supergravity solutions obtained in the preceding sections are compelling candidates for gravity duals to the five-dimensional SCFTs associated with $(p,q)$ five-brane webs. The arguments are based on the following observations. 
\begin{itemize}
\itemsep=0in
 \item[(a)] The solutions have 16 residual supersymmetries and are invariant under $SO(2,5) \oplus SO(3)$.
 \item[(b)] The 3-form field carries non-trivial NS-NS and R-R charges $(p_k,q_k)$ at the poles $r_k$ of a given solution  which are governed by the discontinuity of $\cC$ across~$r_k$. The sum of all the charges is conserved.
 \item[(c)] The behavior of all supergravity fields near each pole $r_k$ matches that of the well-known $(p,q)$ five-branes for the charge assignment $(p_k,q_k)$. Thus, the poles $r_k$ specify the locations of the semi-infinite external five-branes of the web with those charges. 
 \item[(d)] The minimal number of poles required for a non-trivial solution is $3$, matching the minimal number of 
            external semi-infinite $(p,q)$ branes needed to realize an intersection and a five-dimensional SCFT.
 \item[(e)] The number of free parameters in a solution with $L$ poles  is $2L-2$ and matches exactly the number of free parameters of a five-brane intersection with $L$ external branes, given by the external charges subject to overall charge conservation.
\end{itemize}

To examine the solution near a pole $r_k$, we set $w=r_k+\tau  e^{i\theta}$ for $0 \leq \theta \leq \pi$ and expand for small $\tau >0$. The metric function $f_6$ diverges   as $\tau \to 0$ which makes $AdS_6$ expand to Minkowski space-time $\RR^{1,5}$ while $S^2$ combines with the $\theta$-coordinate of $\Sigma$ to form a smooth $S^3$. In this limit, the string-frame metric near $r_k$ becomes
\begin{align}
ds^2&=
ds^2_{\mathds{R}^{1,5}}+
\frac{2}{3} \Big | Z_+^k -Z_-^k \Big | \,\left  (
\frac{d\tau ^2}{\tau ^2}+ds^2_{\mathrm{S}^3} \right ) ~.
\end{align}
The metric  is geodesically complete and regular. The dilaton grows logarithmically with $\tau$, precisely as is expected near a $(p,q)$ five-brane. The entire solution matches precisely onto the near-brane expansion of the $(p,q)$ five-brane solutions  constructed in \cite{Lu:1998vh}, with the following identifications of the $(p,q)$ charges 
\begin{align}
p_k&=\frac{8}{3} \, \Re(Z_+^k)~,
&
q_k&=-\frac{8}{3}\, \Im(Z_+^k)~.
\end{align}
The match includes overall scaling, the form of the geometry and a precise correspondence of overall coefficients.
The constraint $\sum_k Z_\pm^k=0$ implements the charge conservation constraint of the five-brane web.

\sm

A minimal number of three poles is required, since otherwise $\lambda$ has no zeros in the upper half plane and the solution is singular. Since five-dimensional SCFTs are realized on intersections of five-branes,  the geometries obtained by zooming in on the intersection inevitably maintain signatures of the semi-infinite external branes.  This is different from e.g.\ the  case of $\mathcal{N}\,{=}\,4$ supersymmetric Yang-Mills, and its $AdS_5\times S^5$ dual, and is reflected in the presence of the poles.
Solutions with more than 3 poles are possible, and correspond to brane webs with more than 3 external branes.
Gauge theory descriptions of SCFTs corresponding to 3 external branes are less straightforward, 
but have been discussed in \cite{Bergman:2014kza}.

\sm

The remaining free parameters for a solution with $L$ poles are the complex constants $\sigma$ and $\cA^0$, 
the locations of the zeros $s_1,\dots, s_{L-2}$ and the positions of the poles on the real axis $r_1,\dots,r_L$. They make for a total of $3L$ real parameters, which have to satisfy the $L-1$ relations in (\ref{eqn:constr}). Moreover, $SL(2,\RR)$ automorphisms of the upper half plane map to physically equivalent supergravity solutions, further reducing the number of parameters by~$3$. This leaves a total of $2L-2$ real parameters, corresponding precisely to the charge assignments for external  five-branes in the brane web diagrams subject to two real conservation conditions on the charges $(p_k,q_k)$.

\vspace*{4mm}

\section{Discussion}
\label{sec:discussion}

In this letter we have presented the construction of explicit solutions to Type IIB supergravity invariant under the exceptional Lie superalgebra $F(4)$ on a space-time of the form $AdS_6 \times S^2$ warped over a Riemann surface $\Sigma$ with the topology of the upper half plane. We have argued that these solutions represent fully back-reacted and localized intersections of five-branes, and we have exhibited the precise match of their asymptotic behavior with $(p,q)$ five-branes. These solutions have no asymptotically enhanced supersymmetry, contrarily to the more familiar counterparts corresponding to field theory  dimensions 3, 4, and 6, as is consistent with the uniqueness of the superconformal algebra in five dimensions \cite{D'Hoker:2008ix}. They share a remarkable similarity in structure to string-junction solutions with 8 residual supersymmetries to Type 4b supergravity in six dimensions \cite{Chiodaroli:2011nr}. The detailed derivations of the construction will be relayed to a longer paper  \cite{
ads6-2} in which solutions will be considered when $\Sigma$ has more complicated topology, including genus greater than zero, and more than one boundary component. 

\sm

The solutions constructed here are compelling candidates for the holographic duals to the five-dimensional superconformal field theories arising as UV fixed points of non-renormalizable
Yang-Mills theories, and give further support to their existence. Theories on the Coulomb branch and relevant deformations may in principle be constructed and studied. 

\sm

Finally, our solutions provide a stepping stone for direct quantitative analyses, including the calculation of entanglement entropies, free energies, the spectrum of operator dimensions, and correlation functions, issues which we plan to investigate in future work.

\begin{acknowledgments}
We are very happy to thank Andreas Karch for collaboration on earlier work and many illuminating discussions.
We also thank Oren Bergman and David Gieseker for very helpful discussions. The work of all three authors  is supported in part by the National Science Foundation under grant PHY-13-13986 and PHY-16-19926.  
\end{acknowledgments}


%

\end{document}